\documentstyle[aps,floats]{revtex}
\input{psfig.tex}
\begin{document}
\draft
% \twocolumn[\hsize\textwidth\columnwidth\hsize\csname @twocolumnfalse\endcsname
\preprint{\vbox{\hbox{CU-TP-833} 
                \hbox{CAL-635}
%                \hbox{CfA-????}
                \hbox{astro-ph/9705219}
}}

\title{Detectability of Inflationary Gravitational Waves with
       Microwave Background Polarization}

\author{Marc Kamionkowski\footnote{kamion@phys.columbia.edu}}
\address{Department of Physics, Columbia University, 538 West
120th Street,
New York, New York~~10027}
\author{Arthur Kosowsky\footnote{akosowsky@cfa.harvard.edu}}
\address{Harvard-Smithsonian Center for Astrophysics,
60 Garden Street, Cambridge, Massachusetts~~02138
\\and\\
Department of Physics, Lyman Laboratory, Harvard University,
Cambridge, Massachusetts~~02138}
\date{May 1997}
\maketitle

\begin{abstract}
Inflation predicts specific relations between the
amplitudes and spectral indices of the primordial spectrum of
density (scalar metric) perturbations and gravitational waves
(tensor metric perturbations).  Detection of a
stochastic gravitational-wave background is essential for
identifying this
unmistakable signature.  Polarization of the cosmic microwave
background can isolate these tensor
modes in a model-independent way and thereby circumvent the
cosmic-variance limit to detecting a small tensor signal with
only a temperature map.  Here we assess the detectability of a
gravity-wave background with a
temperature and polarization map.  For detector
sensitivities better than $10-20\,\mu$K~$\sqrt{\rm sec}$, the
sensitivity to a tensor signal is always dominated by the polarization
map.  With a detector sensitivity of order $1\,\mu$K~$\sqrt{\rm
sec}$, polarization could improve on a temperature-map
sensitivity to tensor modes by two to three orders of magnitude.
Even a small amount of reionization
substantially enhances tensor-mode detectability.
We also argue that the sensitivity of the Planck Surveyor to
tensor modes is significantly improved with polarization, even
taking into account the resulting degradation of the intensity
determination in the high-frequency channels.

\end{abstract}

\pacs{98.70.V, 98.80.C \hspace*{0.5cm} CU-TP-833, CAL-635,
astro-ph/9705219}
% ]

\def\hatn{{\bf \hat n}}
\def\hatnprime{{\bf \hat n'}}
\def\hatnone{{\bf \hat n}_1}
\def\hatntwo{{\bf \hat n}_2}
\def\hatni{{\bf \hat n}_i}
\def\hatnj{{\bf \hat n}_j}
\def\vecx{{\bf x}}
\def\veck{{\bf k}}
\def\hatx{{\bf \hat x}}
\def\hatk{{\bf \hat k}}
\def\hatz{{\bf \hat z}}
\def\VEV#1{{\left\langle #1 \right\rangle}}
\def\cP{{\cal P}}
\def\noise{{\rm noise}}
\def\pix{{\rm pix}}
\def\map{{\rm map}}
\long\def\comment#1{}

\section{INTRODUCTION}

Slow-roll inflation provides a unified and testable paradigm for
understanding the flatness and smoothness of the Universe and
the origin of structure.  The predictions of a flat Universe
\cite{inflation,newinflation} and a nearly scale-invariant
spectrum of primordial density perturbations (scalar metric
perturbations) \cite{scalars} will be tested with unprecedented
precision by forthcoming cosmic microwave background (CMB)
temperature maps \cite{kamspergelsug,parameters,bet} from 
experiments such as MAP
\cite{MAP} and the Planck Surveyor \cite{PLANCK}.  Inflation
also predicts a nearly scale-invariant stochastic
gravitational-wave background (tensor metric perturbations)
\cite{abbott}. A flat Universe with
nearly scale-invariant adiabatic scalar and tensor perturbations
would certainly suggest that inflation occurred but would
not provide incontrovertible evidence.
However, specific relations between the ``inflationary observables,''  the
amplitudes and spectral indices of the scalar and tensor power
spectra, are unique predictions of inflation \cite{steinhardt}.  
Verification of these relations
would provide a ``smoking-gun'' signature of inflation.  For
this reason, detection of the gravitational-wave background is
of the utmost importance for an unambiguous test of inflation.

Scalar and tensor metric perturbations both contribute to
temperature and polarization fluctuations in the CMB.
A temperature map,
even if ideal, will never determine the tensor amplitude with
a standard error better than around 30\% of the total perturbation
amplitude: cosmic variance from the dominant
scalar perturbations provides a fundamental limit to the sensitivity
of CMB temperature maps to tensor perturbations.  
However, the scalar and tensor contributions to CMB
polarization can be geometrically decomposed in a
model-independent fashion, so one can
circumvent the cosmic-variance limit present in temperature
maps \cite{ourletter,szletter}.  The polarization tensor
$P_{ab}(\hatn)$ on the celestial
sphere can be decomposed into its ``gradient'' (or
``curl-free'') and ``curl'' parts, in analogy to the
decomposition of a vector field.  Scalar perturbations have no
handedness, so they cannot give rise to a curl component.  On
the other hand, tensor perturbations {\it do} have a handedness, so they
induce a curl component.   
The cross-correlation between temperature
and polarization can also help isolate the
tensor contribution, although this signature 
may have some model dependence \cite{cct}.

In this paper, we assess the detectability of inflationary
tensor modes given a full-sky polarization and/or temperature
map with a given angular resolution and level of instrumental
noise.  In other words, for what tensor amplitudes can one hope
to make an unambiguous statistically significant detection of
tensor modes with a given experiment?
We answer this question in two ways:  First, we determine the
minimum tensor amplitude which would be observationally accessible
with only the curl component in the polarization
field.  This will provide a conservative, yet model-independent,
estimate for the sensitivity.  We then consider the tensor
amplitude which would be detectable by fitting all the
temperature and polarization auto- and cross-correlation
functions to inflationary predictions. 

For experiments with detector sensitivities near
$10\,\mu$K~$\sqrt{\rm sec}$, the tensor-mode detectability with a
polarization map can improve on that from a temperature map alone
by an order of magnitude. Even with detectors substantially
less sensitive, polarization can provide large improvements in
tensor detectability for cases where the temperature map is
not very restrictive, as when fitting many cosmological parameters
to a single data set.  In this case, the temperature-polarization
cross-correlation, which has a larger amplitude than the curl
polarization component, becomes important. Detector sensitivities
attainable in the foreseeable future---say 
$1\,\mu$K~$\sqrt{\rm sec}$---will dramatically enhance our
ability to see tensor perturbations; tensor-scalar ratios
as small as $10^{-4}$ might be probed.  Even a small amount of
reionization significantly improves the detectability of
tensor modes.

Below, we begin with a review of the inflationary observables
in Section II.  Section III reviews the gradient/curl
decomposition and the temperature-polarization power spectra.
In Section IV, we calculate the model-independent tensor
sensitivity achievable with measurements of only the curl
component of the polarization.  We then consider what can be
learned from only a temperature map and from a combined
temperature-polarization map.  Section V compares these
sensitivities with the predictions of some specific inflationary
models, and some concluding remarks follow in Section VI.

\section{THE INFLATIONARY OBSERVABLES}

Inflation occurs when some scalar field $\phi$ (the ``inflaton'')
is displaced from the minimum of its potential $V(\phi)$ such
that the energy density of the Universe is dominated by the field's 
potential energy for a time long compared with the Hubble time.
During this inflationary phase, the expansion of the Universe
accelerates, small quantum fluctuations
in $\phi$ produce classical scalar perturbations, and quantum
fluctuations in the spacetime metric produce gravitational waves.
If the inflaton potential $V(\phi)$ is given in units
of $m_{\rm Pl}^4$, and the inflaton $\phi$ is in units of $m_{\rm Pl}$,
then the scalar and tensor spectral indices are
\begin{eqnarray}
     1-n_s &=& { 1 \over 8\pi} \left( {V' \over V} \right)^2 -
     {1 \over 4 \pi} \left({V' \over V} \right)', \nonumber\\
     n_t &=& -{ 1 \over 8\pi} \left( {V' \over V} \right)^2. 
\label{spectralindices}
\end{eqnarray}
The amplitudes can be fixed by their contribution to $C_2^{\rm TT}$,
the quadrupole moment of the CMB temperature,
\begin{eqnarray}
     {\cal S} &\equiv & 6\, C_2^{{\rm TT},{\rm scalar}}= 33.2\,[V^3/(V')^2],
          \nonumber\\
     {\cal T} &\equiv & 6\, C_2^{{\rm TT},{\rm tensor}}= 9.2 \,V.
\label{amplitudes}
\end{eqnarray}
For the slow-roll conditions to be satisfied, we must have
\begin{eqnarray}
     (1 /16 \pi) (V'/V)^2 &\ll& 1, \\ 
     (1 /8\pi)(V''/V) & \ll & 1,
\label{slowrollconditions}
\end{eqnarray}
which guarantee that inflation lasts long enough to make the Universe
flat and to solve the horizon problem.

When combined with COBE results, current degree-scale anisotropy and
large-scale structure observations suggest that ${\cal T}/{\cal S}$ is less
than order unity in inflationary models, which restricts
$V\lesssim 5\times 10^{-12}$.  If the consistency relation
${\cal T} / {\cal S} \simeq -7 n_t$ [implied by
Eqs. (\ref{spectralindices}) and (\ref{amplitudes})] holds, the
tensor spectrum must be nearly scale invariant ($n_t\simeq 0$).

\section{STATISTICS OF CMB ANISOTROPIES AND POLARIZATION}

\subsection{Harmonic Expansion}

A temperature map $T(\hatn)$ of the sky can be expanded in
spherical harmonics,
\begin{equation}
     {T(\hatn) \over T_0} = 1 + \sum_{lm} a^{\rm T}_{(lm)}
     Y_{(lm)}(\hatn),
\label{Texpansion}
\end{equation}
where the mode amplitudes are given by
\begin{equation}
     a^{\rm T}_{(lm)}={1\over T_0}\int
     d\hatn\,T(\hatn)\,Y_{(lm)}^*(\hatn);
\label{Talms}
\end{equation}
this follows from the orthonormality of the spherical harmonics.

The Stokes parameters $Q(\hatn)$ and $U(\hatn)$
(where $Q$ and $U$ are measured with respect to the polar ${\bf
\hat\theta}$ and azimuthal ${\bf \hat \phi}$ axes) which specify
the linear polarization in direction $\hatn$ are components of a
$2\times2$ symmetric trace-free (STF) tensor, 
\begin{equation}
  {\cal P}_{ab}(\hatn)={1\over 2} \left( \begin{array}{cc}
   \vphantom{1\over 2}Q(\hatn) & -U(\hatn) \sin\theta \\
   -U(\hatn)\sin\theta & -Q(\hatn)\sin^2\theta \\
   \end{array} \right),
\label{whatPis}
\end{equation}
where the subscripts $ab$ are tensor indices.
Just as the temperature is expanded in terms of spherical
harmonics, the polarization tensor can be expanded,
\cite{ourlongpaper}
\begin{equation}
      {{\cal P}_{ab}(\hatn)\over T_0} =
      \sum_{lm} \left[ a_{(lm)}^{{\rm G}}Y_{(lm)ab}^{{\rm
      G}}(\hatn) +a_{(lm)}^{{\rm C}}Y_{(lm)ab}^{{\rm C}}(\hatn)
      \right],
\label{Pexpansion}
\end{equation}
in terms of the tensor spherical harmonics $Y_{(lm)ab}^{\rm G}$
and $Y_{(lm)ab}^{\rm C}$, which are a complete basis for the
``gradient'' (i.e., curl-free) and ``curl'' components of the
tensor field, respectively. (See Ref. \cite{szlongpaper} for an
alternative but equivalent formalism.) The mode amplitudes are given by
\begin{eqnarray}
a^{\rm G}_{(lm)}&=&{1\over T_0}\int d\hatn\,{\cal P}_{ab}(\hatn)\, 
                                         Y_{(lm)}^{{\rm G}
					 \,ab\, *}(\hatn),\cr 
a^{\rm C}_{(lm)}&=&{1\over T_0}\int d\hatn\,{\cal P}_{ab}(\hatn)\,
                                          Y_{(lm)}^{{\rm C} \,
					  ab\, *}(\hatn), 
\label{Amplitudes}
\end{eqnarray}
which can be derived from the orthonormality properties,
\begin{eqnarray}
\int d\hatn\,Y_{(lm)ab}^{{\rm G}\,*}(\hatn)
             Y_{(l'm')}^{{\rm
	     G}\,\,ab}(\hatn)&=&\delta_{ll'}\delta_{mm'},  \cr 
\int d\hatn\,Y_{(lm)ab}^{{\rm C}\,*}(\hatn)
             Y_{(l'm')}^{{\rm
	     C}\,\,ab}(\hatn)&=&\delta_{ll'}\delta_{mm'},  \cr 
\int d\hatn\,Y_{(lm)ab}^{{\rm G}\,*}(\hatn)
             Y_{(l'm')}^{{\rm C}\,\,ab}(\hatn)&=&0.
\label{Orthonormality}
\end{eqnarray}
Here $T_0$ is the cosmological mean CMB temperature and we are
assuming $Q$ and $U$ are measured in brightness temperature
units rather than flux units.   Scalar perturbations have no
handedness.  Therefore, they can produce no curl, so
$a_{(lm)}^{\rm C}=0$ for scalar modes.  On the other hand tensor modes
{\it do} have a handedness, so they produce a
non-zero curl, $a_{(lm)}^{\rm C} \neq0$.

\subsection{Statistics}

A given inflationary model predicts that the $a_{(lm)}^{\rm X}$
are distributed with zero mean, $\VEV{a_{(lm)}^{\rm X}}=0$ (for $\{
{\rm X} = {\rm T,G,C}\}$), and covariance
$\VEV{\left(a_{(l'm')}^{\rm X'} \right)^* a_{(lm)}^{\rm X}} =
C_l^{{\rm XX}'} \delta_{ll'}\delta_{mm'}$.  Parity demands that
$C_l^{\rm TC}=C_l^{\rm GC}=0$.  Therefore the statistics of the
CMB temperature-polarization map are completely specified by the
four sets of moments, $C_l^{\rm TT}$, $C_l^{\rm TG}$, $C_l^{\rm
GG}$, and $C_l^{\rm CC}$.   Also, as stated above, only tensor modes
will produce nonzero $C_l^{\rm CC}$.  

\begin{figure}[htbp]
\centerline{\psfig{file=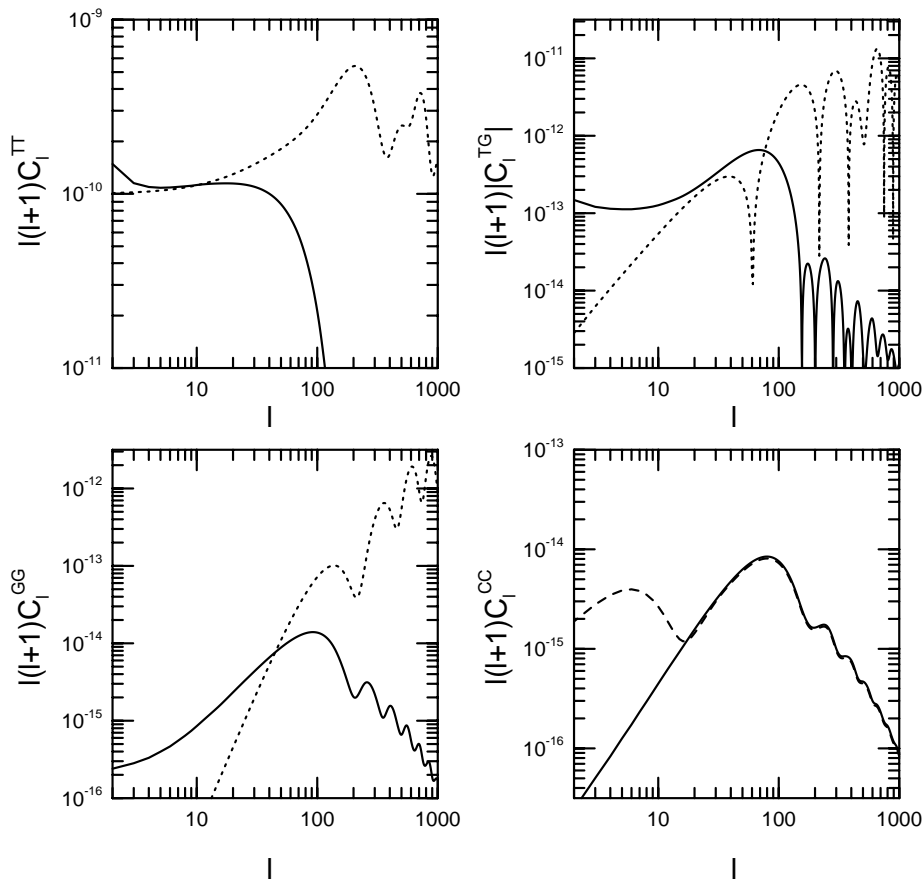,width=7in}}
\bigskip
\caption{
          Theoretical predictions for the four nonzero CMB
	  temperature-polarization spectra as a function
	  of multipole moment $l$.  The solid curves are the
	  predictions for a COBE-normalized inflationary model
	  with no reionization and no gravitational waves for
	  $h=0.65$, $\Omega_b
	  h^2=0.024$, and $\Lambda=0$.  The dotted curves are the
	  predictions which would be obtained if the COBE
	  anisotropy were due entirely to a stochastic
	  gravity-wave background with a flat scale-invariant
	  spectrum (with the same cosmological parameters).
	  Note that the panel for $C_l^{\rm CC}$ 
          contains no dotted curve since scalar perturbations
	  produce no ``C'' polarization component; instead,
          the dashed line in the lower right panel shows a
	  reionized model with optical depth $\tau=0.1$ to the
	  surface of last scatter.
       }
\label{clsplot}
\end{figure}

To illustrate, Fig.~\ref{clsplot} shows the four
temperature-polarization power spectra. The dotted curves correspond
to a COBE-normalized inflationary model
with cold dark matter and no cosmological constant
($\Lambda=0$), Hubble constant (in units of 100~$h$ km~sec$^{-1}$~Mpc$^{-1}$)
$h=0.65$, baryon density $\Omega_bh^2=0.024$, scalar spectral
index $n_s=1$, no reionization, and 
no gravitational waves.  The solid curves show the spectra for
a COBE-normalized stochastic gravity-wave background with a
flat scale-invariant spectrum ($h=0.65$, $\Omega_b h^2=0.024$,
and $\Lambda=0$) in a critical-density Universe.  
Note that the panel for $C_l^{\rm CC}$ contains no dotted curve
since scalar perturbations produce no C polarization component.
The dashed curve in the CC panel shows the tensor spectrum
for a reionized model with optical depth $\tau=0.1$ to the
surface of last scatter.  We use the code CMBFAST
\cite{szlongpaper} to generate all power spectra used in this paper
(but note that the normalizations of the polarization $C_l$'s
{}from this code are different than those used in this paper
\cite{ourlongpaper}).

\subsection{Cosmic and Pixel-Noise Variance}

Theory predicts the covariances $C_l^{{\rm XX}'}$ of the
distributions for the
multipole coefficients $a_{(lm)}^{\rm X}$.  
For a given $l$, only $2l+1$ independent harmonics $(m= -l,\ldots ,l)$
are available to estimate these variances, giving a statistical
``cosmic variance'' limit
to how well we can estimate each moment $C_l^{{\rm XX}'}$ from a map.
Instrumental noise, beam size, and sky coverage also contribute
to the variance in each moment.
If $\widehat{C_l^{{\rm XX}'}}$ is an
estimator for the moment $C_l^{\rm XX'}$, 
then the ($6\times6$) covariance matrix for these estimators is
\begin{equation}
     \Xi_{{\rm AB}} \equiv \VEV{ \left( \widehat{C_l^{\rm
     A}} - C_l^{\rm A} \right)
     \left( \widehat{C_l^{{\rm B}}} - C_l^{{\rm B}} \right) },
\label{Xidefn}
\end{equation}
for A,B$={\rm XX}'$.  Expressions for the entries of this matrix
for an idealized mapping experiment with a given instrumental noise,
beamwidth, and fraction of sky covered are presented in Section III
of Ref. \cite{ourlongpaper}.

\section{DETECTABILITY OF TENSOR MODES}

\subsection{Curl Component of the Polarization}

The curl component of polarization, $C_l^{\rm CC}$,
provides a model-independent probe of tensor
perturbations in inflationary models.  
What detector sensitivity is required to distinguish this signal
{}from a null result?  The answer to
this question will of course depend on the angular spectrum of
the curl component, or equivalently, the $l$ dependence of
$C_l^{\rm CC}$.  
However, as noted above, current measurements
indicate that $n_t$ must be close to zero if the inflationary
relation ${\cal T}/{\cal S} \simeq -7n_t$ is satisfied.
Therefore, in the following analysis we consider a scale-invariant
($n_t=0$) tensor perturbation spectrum.  
Furthermore, variations in the other
cosmological parameters have a relatively weak effect on the
spectrum of the curl component (except for reionization, the effect
of which is mentioned below), and it is likely these
parameters will be fairly well determined by temperature
maps and other observations.

Consider a mapping experiment which measures the
temperature and polarization on the entire sky with
beamwidth $\theta_{\rm FWHM}$ and a temperature sensitivity
$s$ (which has units $\mu$K~$\sqrt{\rm sec}$),
giving $N_{\rm pix}\simeq 42,000\,(\theta_{\rm FWHM}/{\rm
deg})^{-2}$ independent pixels on the sky.
The gaussian beamwidth is $\sigma_b\equiv 7.42\times 10^{-3}
(\theta_{\rm FWHM}/1^\circ)$.
If the Universe has no tensor perturbations, then the $1\sigma$ upper
limit to the tensor amplitude in a null
experiment is $\sigma_{\cal T}$, where
\begin{equation}
     {1\over \sigma_{\cal T}^2} = \sum_l \left( { \partial C_l^{\rm CC}
     \over \partial {\cal T}} \right)_{{\cal T}=0}^2 {1\over
     (\sigma_l^{\rm CC})^2},
\label{simplest}
\end{equation}
and 
\begin{equation}
     \sigma_l^{\rm CC} = \sqrt{2/(2l+1)} w^{-1} e^{l^2
     \sigma_b^2},
\label{CCnoise}
\end{equation}
is the pixel-noise variance with which $C_l^{\rm CC}$ can be
determined (there is no cosmic variance if there is no
cosmological signal).  Here, $w^{-1}=4\pi s^2 /(t_{\rm pix}
N_{\rm pix}T_0^2)$ is the inverse weight per unit area on the sky
where $t_{\rm pix}$ is the time spent observing
each pixel, so $w^{-1}=2.14\times10^{-15} t_{\rm yr}^{-1}(s/200\, \mu{\rm K}\,
\sqrt{\rm sec})^2$ with $t_{\rm yr}$ the total observing
time in years.  Since $\cal T$ is the overall normalization of
the tensor spectrum, we can write $\partial C_l^{\rm
CC}/\partial {\cal T} = C_l^{\rm CC}/{\cal T}$. 
Substituting the $C_l^{\rm CC}$ spectrum from Fig.~1 (no reionization)
into Eq.~(\ref{simplest}) with $\theta_{\rm FWHM}=0.5$ gives
\begin{equation}
     {\sigma_{\cal T}\over 6\, C_2^{\rm TT}}
      \simeq 5\times 10^{-4} \left( {s\over \mu{\rm K}\,\sqrt{\rm
      sec}} \right)^2 t_{\rm yr}^{-1}.
\label{CCresult}
\end{equation}
Thus, the curl component
of a full-sky polarization map is sensitive to inflaton
potentials $V\gtrsim 5 \times 10^{-15}t_{\rm yr}^{-1}$ $(s/\mu{\rm K}\,
\sqrt{\rm sec})^2$.  Tensor modes produce
polarization primarily on angular scales greater than a degree;
the result in Eq.~(\ref{CCresult}) will be similar
for any $\theta_{\rm FWHM} \lesssim 1^\circ$.  Improvement on
current constraints with only the curl
polarization component requires a detector
sensitivity $s\lesssim40\,t_{\rm yr}^{1/2}\,\mu$K$\sqrt{\rm
sec}$. Again, the curl component of polarization
is due only to tensor perturbations and its
shape is insensitive to the baryon density and Hubble
constant.  

Even a small amount of reionization will significantly increase
the polarization signal at low $l$, 
as shown in the CC panel of Fig.~1, making the above
result sensitive to the ionization history of the Universe.
For example, in the same
model with an optical depth to last scattering of $\tau=0.1$,
the numerical factor in Eq.~(\ref{CCresult}) becomes $9.3\times 10^{-5}$, 
increasing sensitivity to the tensor
modes by more than a factor of 5.
This level of reionization (if not more) is expected
in cold dark matter models \cite{kamspergelsug,blanchard,haiman},
so Eq.~(\ref{CCresult}), for no reionization, 
provides a conservative estimate.

\subsection{Full Polarization and Temperature Spectra}

Fitting an inflationary model to the entire set of temperature
and polarization power spectra can improve tensor detectability,
especially at comparatively low sensitivities.
Generalizing Eq.~(\ref{simplest}),
if the tensor amplitude is the only relevant parameter,
the $1\sigma$ sensitivity to the tensor amplitude is given by
\begin{equation}
     {1\over \sigma_{\cal T}^2} = \sum_l \sum_{\rm XX'}
     \left({\partial C_l^{\rm X} \over \partial {\cal T}}
     \right)_{{\cal T}=0}
     [\Xi^{-1}]_{\rm XX'} \left({\partial C_l^{{\rm X}'} 
            \over \partial {\cal T}}
     \right)_{{\cal T}=0},
\label{notsimplest}
\end{equation}
where the second sum is over ${\rm X,X}' = \{{\rm TT,TG,GG,CC} \}$.
However, the TT, TG, and GG power spectra also have a strong
dependence on other cosmological
parameters.  Furthermore, some parameters, such as the
scalar-mode normalization and power-law index, can only be determined with any
precision from the CMB.  Multiple parameters can be
accounted for with the
curvature matrix (also known as the Fisher information matrix)
\begin{equation}
     \alpha_{ij} = \sum_l \sum_{\rm XX'} {\partial C_l^{\rm
     X}({\bf s_0}) \over 
     \partial s_i} [\Xi^{-1}]_{\rm XX'} {\partial C_l^{{\rm
     X}'}({\bf s_0}) \over 
     \partial s_j} + P_{ij},
\label{fullvariance}
\end{equation}
where ${\bf s_0}$ are the parameters of the putative underlying
cosmological model.  The matrix $P_{ij}$ takes into account
gaussian priors for these parameters ${\bf s}_0$ \cite{bet}; it
is the inverse of the covariance matrix for the parameters 
determined from other
measurements or observations. 
The covariance matrix is the
inverse of the curvature matrix, ${\cal C} = [\alpha]^{-1}$,
and the standard error in the tensor amplitude, after marginalizing
over all other undetermined parameters, is $\sigma_{\cal T}=\sqrt{{\cal
C}_{\cal TT}}$.  Inclusion of the prior matrix $P_{ij}$ guarantees
that the other undetermined parameters will be marginalized over
only for reasonable values of those parameters.  If it is not
included, the results for $\sigma_{\cal T}$ are overly
conservative.

Note that when testing the null hypothesis of no tensor
perturbations, the underlying model $\bf s_0$ has no tensor
contribution, so the derivatives of $C_l^{\rm CC}$ in Eq.~(\ref{fullvariance})
with respect to all the parameters except $\cal T$ are zero. In
this case, the CC contribution to the variance decouples from
the rest of the power spectra and has no dependence on the other
cosmological parameters. However, even in this case, the other
three power spectra depend on all the cosmological parameters,
so the full curvature matrix must be calculated when comparing to
the complete set of temperature and polarization data. 

\begin{figure}[htbp]
\centerline{\psfig{file=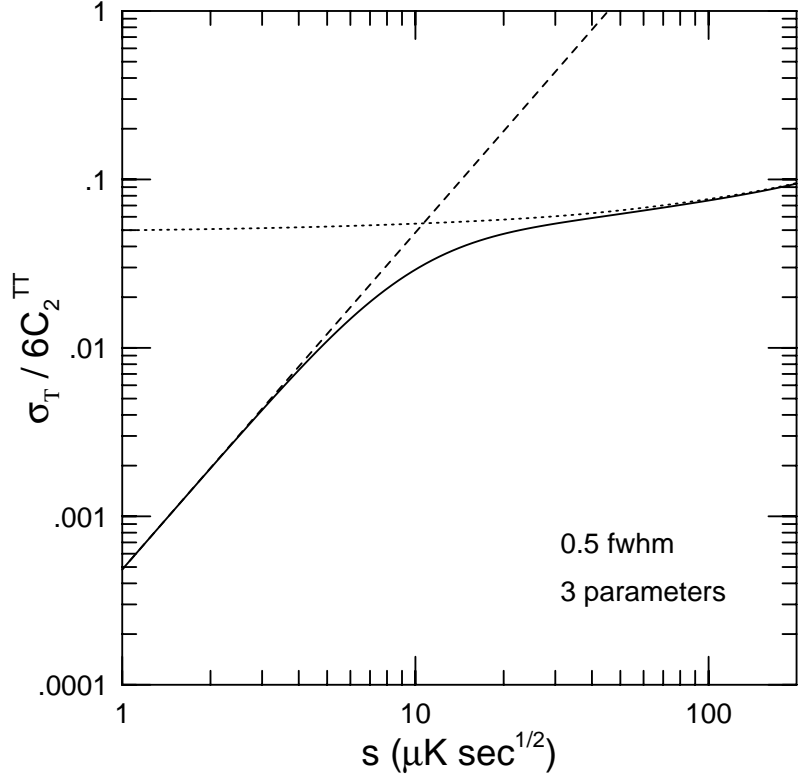,width=6in}}
\caption{
         Results for the $1\sigma$ sensitivity $\sigma_{\cal T}$ to
	 the amplitude $\cal T$ of a flat ($n_t=0$) tensor spectrum
	 as a function of detector sensitivity $s$ for an
	 experiment which maps the CMB temperature and
	 polarization on the full sky for one year with an
	 angular resolution of $0.5^\circ$.  The vertical axis is in
         units of the temperature quadrupole. Here, we have
	 assumed that the spectra are fit only to $\cal S$,
	 $\cal T$, and $n_s$, and the parameters of the
	 cosmological model are those used in Fig.~1.  The dotted curve
	 shows the results obtained by fitting only the TT
	 moments; the dashed curve shows results obtained by
	 fitting only the CC moments; and the solid curve shows
	 results obtained by fitting all four sets of moments. 
}
\label{resultssimple}
\end{figure}

\begin{figure}[htbp]
\centerline{\psfig{file=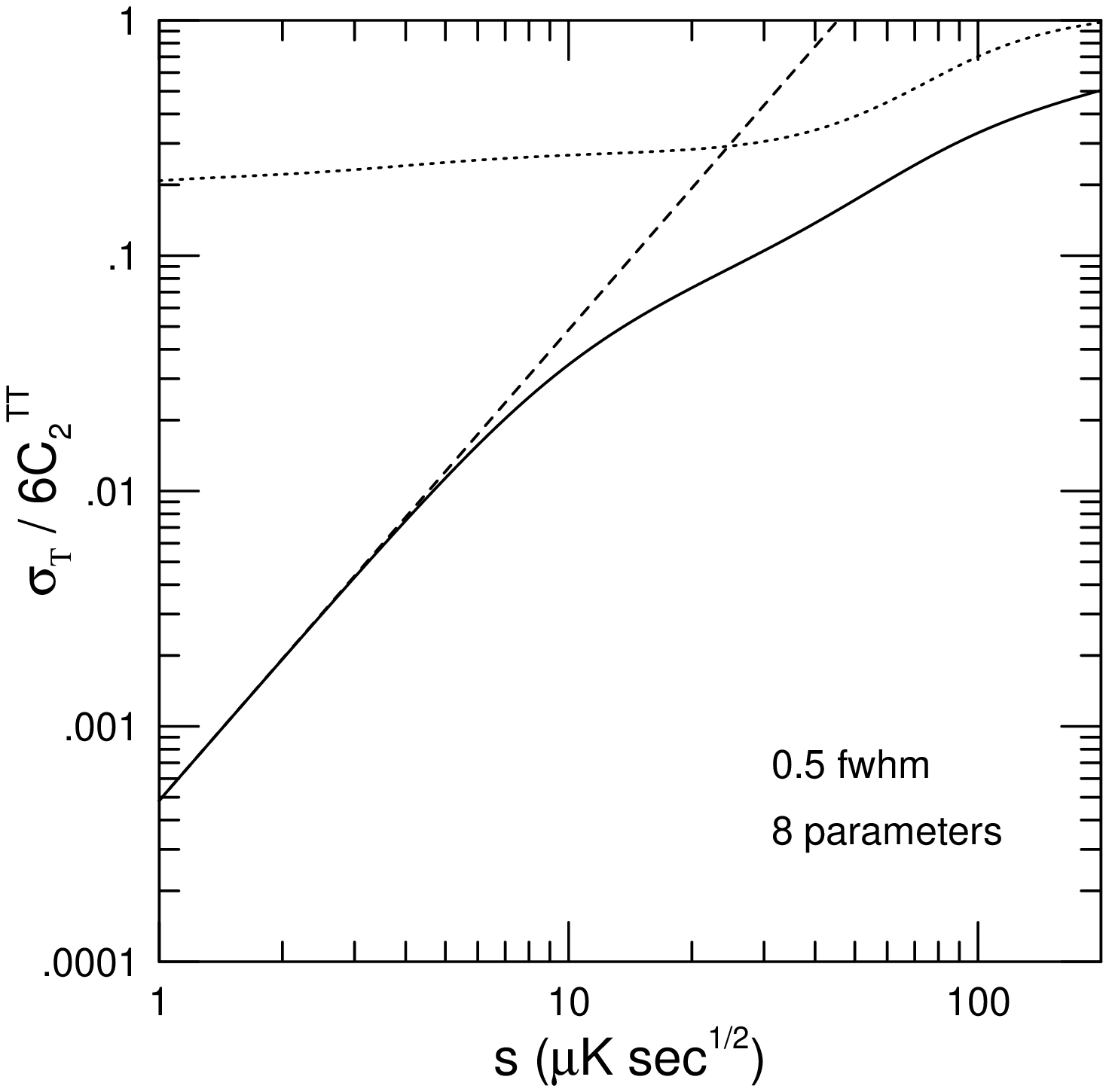,width=6in}}
\caption{
          The same as in Fig.~2, but here we assume that we will
	  be fitting for $h$, $\Omega_b h^2$, $\Lambda$, $\tau$,
	  and $\Omega_\nu h^2$ (with very conservative priors),
	  in addition to $\cal S$, $\cal T$, and $n_s$.
}
\label{resultscomplicated}
\end{figure}

Figures \ref{resultssimple} and \ref{resultscomplicated}
show the results for $\sigma_{\cal T} = \sqrt{{\cal C}_{\cal TT}}$, obtained
{}from Eq.~(\ref{fullvariance}), as a function of detector
sensitivity $s$ for a full-sky, one-year mapping experiment.  We
have used a beamwidth $0.5^\circ$, but the result is essentially
independent of beam size for any beam smaller than one degree. 
In both cases, the parameters of the putative underlying model
are those used in Fig.~1, and we again consider the
sensitivity to a tensor spectrum with $n_t=0$.  
In both Figures, the solid curve shows
the results which would be obtained by fitting all four sets of
temperature and polarization moments; the dotted line shows
results which would be obtained by fitting only the temperature
moments, while the dashed line shows those obtained by fitting
only the CC polarization.

In Figure \ref{resultssimple}, we make the optimistic
assumption that all cosmological parameters except the scalar
normalization $\cal S$ and spectral index $n_s$ are known.
Therefore, we diagonalize the $3\times3$ covariance matrix 
for $\cal T$, $\cal S$, and
$n_s$, assuming no prior information
about any of these quantities.  
Figure \ref{resultscomplicated} is
more conservative.  Here we assume that $h$, $\Omega_b
h^2$, $\Lambda$, the massive-neutrino density $\Omega_\nu h^2$,
and $\tau$ are to be determined from the data as well as $\cal S$,
$\cal T$, and $n_s$.  We include only very conservative priors
on these parameters:  $1\sigma$ gaussian
errors of 0.01 on $\Omega_b h^2$, 0.3 on $\Lambda$, 0.15 on $h$,
0.5 on $\tau$, and 0.2 on $\Omega_{\nu} h^2$.

For detectors sensitivities $s\gtrsim 20\,\mu$K$\,
\sqrt{\rm sec}$, the tensor-mode detectability with a
three-parameter fit comes primarily
{}from the temperature map, although polarization does provides
some incremental improvement.  However, for detector
sensitivities $s \lesssim10\,\mu$K$\sqrt{\rm sec}$, the
sensitivity to tensor modes comes almost entirely from the
polarization. For the eight-parameter fit (which
is probably closer to the types of fits which will be
done with satellite data), the determination
of the tensor amplitude is always dominated by polarization, although
in this case the temperature-polarization cross-correlation gives
most of the contribution at larger values of $s$.
Note that the solid curves (from fitting to
the complete temperature and polarization power spectra) in
Figs.~\ref{resultssimple} and \ref{resultscomplicated} asymptote
to the dotted curves (from fitting only to the CC spectrum) at
smaller $s$. This indicates that for better detector
sensitivities, the determination of the tensor amplitude comes primarily
{}from the curl component of the polarization and thus is largely
independent of the other cosmological 
parameters.  Any beam size below about a degree will
be small enough to detect virtually all of the CC signal.  Beams
smaller than this will only improve the sensitivity to tensor
modes by better constraining the other cosmological 
parameters from the TT and TG spectra.

In reionized models, the dashed CC curves in Figs.~2 and 3 move down
substantially, while the dotted temperature curves remain essentially
unchanged. As mentioned above, an optical depth of $\tau=0.1$ gives a 
factor of 5 improvement in detectability from the CC power spectrum,
bringing $\sigma_{\cal T}/6C_2^{\rm TT}$ below $10^{-4}$ for
the foreseeable detector sensitivity of $1\,\mu$K$\sqrt{\rm sec}$.
Figures \ref{resultssimple} and
\ref{resultscomplicated}, for no reionization, are
conservative estimates of tensor mode detectability.

With traditional bolometer detectors 
(e.g., the high-frequency channels on Planck),
the polarization is determined by placing a polarizing filter
in front of the detector, thereby halving the number of photons
available for the temperature measurement. In this case,
a temperature-only map with a given detector sensitivity $s$
should be compared with a polarized map of sensitivity $\sqrt{2} s$.
Doing so, it is clear that
for any detector sensitivity $s\lesssim 200\, \mu$K$\sqrt{\rm
sec}$ (well above the sensitivity of the Planck high-frequency
channels), the sensitivity to a tensor signal is improved
significantly with the inclusion of polarization, even after
taking into account the degradation of the temperature signal.
Note that for Planck, an rms pixel noise of $\Delta T/T
= 2\times10^{-6}$ with angular resolution of $10'$ \cite{PLANCK} gives
$s\simeq 25\, \mu$K$\sqrt{\rm sec}$.  If the high-frequency
channels are polarized, this number becomes $s\simeq 35\,
\mu$K$\sqrt{\rm sec}$ after including the extra factor of
$\sqrt{2}$.

\section{SOME SPECIFIC MODELS}

It is quite plausible
that the amplitude of an inflationary stochastic gravity-wave
background is large enough to be seen with a next-generation
experiment. Slow-roll inflation
provides a beautiful and economical explanation of isotropy,
flatness, and the origin of density perturbations.  However, we do
not know the details of the physics responsible for inflation.
Some models predict a sizeable tensor amplitude,
while in others it is hopelessly small.  Unfortunately,
no consensus exists among theorists as to which are most likely.
To illustrate the state of theoretical expectations for the
amplitude of tensor modes, this Section give a brief overview of some
inflationary models.  Rather than survey the plethora of
specific models individually, we follow the classification of
Ref.~\cite{dodelson}.  We also suggest
Ref.~\cite{lyth} for an intriguing parallel discussion.

In a large-field polynomial
potential (e.g., chaotic inflation \cite{chaotic}), $V(\phi)
\propto \phi^p$ with $p>1$, the expected tensor amplitude is
${\cal T} \simeq 1.4 \times 10^{-9}\, p/(p+200)$.  This
tensor amplitude will be detectable through the curl polarization
component alone with a sensitivity 
$s\simeq 16\, \mu$K~$\sqrt{\rm sec}$ for $p=2$, for example. 
Therefore,
the tensor modes should be accessible with Planck if this is the
correct model for inflation.

In small-field polynomial potentials, $V(\phi) \propto [ 1 -
(\phi/\mu)^p]$ with $\phi \ll \mu \ll m_{Pl}$ and $p>2$, the
tensor spectrum is unobservably small.  This is the type of
potential expected if inflation occurs from a
spontaneous-symmetry-breaking transition such as those
envisioned in new inflation \cite{newinflation}.
Similarly, in small-field quadratic potentials, $V(\phi) \propto
[1 - (\phi/\mu)^2]$ with $\phi \ll \mu$, the tensor amplitude
is unobservably small.  Such a potential arises, e.g., in
``natural inflation'' models \cite{naturalinflation}.

In a linear potential, $V(\phi) \propto \phi$, the tensor
amplitude is proportional to the deviation of the scalar
amplitude from unity; i.e., ${\cal T}\simeq 7\times 10^{-10}\,
(n_s-1)$. To detect such a signal through the curl polarization
component alone requires a
sensitivity $s\simeq 60 \mu$K~$\sqrt{\rm sec}
\,[(n_s-1)/0.3]^{1/2}$. A similar situation arises in exponential
models, $V(\phi) \propto \exp \sqrt{16 \pi \phi^2 / p m_{Pl}^2}$
with $p>0$.  However, the constant of proportionality differs
slightly.  These models require a sensitivity
$s\simeq 40 \mu$K~$\sqrt{\rm sec} \,[(n_s-1)/0.3]^{1/2}$ to
detect the tensors with only the curl polarization component.  Therefore,
if the scalar tensor index deviates by ${\cal O}(10\%)$---which
is consistent, although not necessarily indicated, by
COBE---this tensor signal can be detected with Planck.  On
the other hand, if $n_s$ is in
fact very close to unity, the tensor signal might turn out to be
unobservably small.
In hybrid inflation models (those which require two fields for
inflation), $V(\phi) \propto [1+(\phi/\mu)^p]$ with $\phi < \mu$
and $p\geq 2$ \cite{hybrid}, the tensor 
amplitude is only constrained to be smaller than that in
exponential models, so
a tensor signal may be observable in such models.  

To summarize, some reasonable inflationary models give
a tensor perturbation signal within reach of 
next-generation polarization experiments, whereas in many
others the signal will be elusive.  As this brief survey
illustrates, CMB polarization can help discriminate
between models.  Detection of tensor modes would dramatically
narrow the available range of models and determine the energy
scale of inflation, while a null result would also provide
interesting constraints on models.

\section{DISCUSSION AND CONCLUSIONS}

Detection of a stochastic gravity-wave background is essential
to test the full predictions of slow-roll inflation.  Here
we have
evaluated the detectability of tensor metric perturbations with a
polarization map of the CMB.  Inclusion of polarization will
always improve on the tensor-mode sensitivity achievable with only a
temperature map.  For detector sensitivities $s \gtrsim
10-20\,\mu$K~$\sqrt{\rm sec}$, the improvement is incremental
and comes primarily from the temperature-polarization
cross-correlation;  for $s \lesssim
10-20\,\mu$K~$\sqrt{\rm sec}$, 
the improvement is dramatic and tensor-mode detectability is dominated by
the polarization, particularly the curl (CC)
component of the polarization.  For detector sensitivities of
$s=1\,\mu$K~$\sqrt{\rm sec}$ the improvement is by two to three
orders of magnitude.  Polarization will significantly enhance the
sensitivity to tensor modes for the
high-frequency channels on Planck, even though the temperature
measurement must be degraded to accommodate polarization.
The tensor signature in the TT, TG, and
GG spectra is somewhat model dependent, but the CC spectrum,
which dominates the tensor signal with better detector
sensitivities, provides an unambiguous model-independent probe of
the stochastic gravity-wave background.

The ability of a temperature map to detect tensor perturbations
is limited by cosmic variance to around 30\% of the total
perturbation amplitude.
Some prior work \cite{knoxturner,mikedetect}
concluded that cosmic variance would also limit the sensitivity of
a polarization map to tensor modes.  However, these papers
did not take into account the geometric decomposition of scalar
and tensor modes which is possible with a polarization map.
Since scalar perturbations make no contribution to the CC
polarization spectrum, the cosmic-variance limitations are
effectively circumvented.
The detectability of tensor modes with terrestrial and
space-based gravitational-wave detectors was addressed in
Ref.~\cite{mikedetect}, although we believe that CMB
polarization provides a more promising avenue toward
detection of the predominantly longer-wavelength inflationary
gravitational radiation. 

We have also reviewed some inflationary models.  
Predictions for the tensor amplitude differ greatly
among plausible models;  
a high-sensitivity polarization map will help
discriminate between inflationary models.  In the event of a
positive detection, the relations between the inflationary
observables can be tested: the relation
${\cal T}/{\cal S} \simeq -7 n_t$ \cite{steinhardt} must be satisfied.
Several authors have previously investigated
how precisely the inflationary observables can be determined with a
temperature map alone \cite{parameters,bet,dodelson,knox}
and with a temperature-polarization map \cite{zss} for some
assumed (perhaps optimistic) models in which the amplitude of
the tensor signal is large enough to be detected.  
While cosmic variance from scalar modes essentially precludes 
a temperature map from determining $n_t$, the CC polarization
spectrum isolates the gravitational waves and thereby
allows determination of $n_t$.  For example,
with ${\cal T}/{\cal S}=0.05$ and a detector
sensitivity of a few $\mu$K$\sqrt{\rm sec}$, $n_t$ may be determined
with a standard error of 0.05 or better by a one-year mapping experiment.

We note briefly that polarization measurements at very high
sensitivities will require projecting out contributions
{}from foreground polarized emission. While the characteristics
and amplitudes of polarization foregrounds are unknown at present, the
valuable cosmological signals contained in CMB polarization
warrant intensive study of this question.

CMB polarization may be used to address a number
of cosmological issues aside from tensor perturbations.  
Polarization of order 5--10\% of the
temperature anisotropy is predicted in any model in which the
CMB has a cosmological origin \cite{be}.  The absence of
polarization or a polarization greatly in excess of that
expected would force serious reconsideration of current cosmological
models at the most fundamental level.  
Polarization can
disentangle the peculiar-velocity contributions
to the anisotropy on smaller angular scales \cite{zaldharari};
only with a combined temperature-polarization map
can a unique and unambiguous reconstruction of primordial density
perturbations be made.  Polarization can provide
incontrovertible evidence for acoustic
oscillations in the early Universe \cite{akunpub},
help distinguish between competing cosmological models
\cite{spt}, constrain the ionization history of the Universe
\cite{reionization}, 
probe cosmological magnetic fields \cite{arthuravi1},
and improve cosmological-parameter
determination \cite{zss}.  The polarization of the CMB
toward clusters can potentially be used to access other
CMB surfaces of last scatter \cite{marcavi} or to learn about
cluster physics \cite{clusters,arthuravi2}.

Detection techniques and study of polarization foregrounds are
in their infancy when compared with temperature anisotropies.
However, we hope that the arguments presented here 
illustrate the fundamental importance of CMB polarization for
physics and cosmology and motivate experimental developments in
this direction.

\acknowledgments

MK thanks George Smoot for useful conversations.  This work was
supported by D.O.E. contract DEFG02-92-ER 40699, NASA contract
NAG5-3091, and the Alfred P. Sloan Foundation at Columbia, and
by the Harvard Society of Fellows.

\end{document}